# Software Conceptual Integrity: Deconstruction, Then Reconstruction

Iaakov Exman

*Software Engineering Department*
*The Jerusalem College of Engineering- Azrieli*
*Jerusalem, Israel*
*iaakov@jce.ac.il*

**Abstract**

*Conceptual Integrity* is the most important consideration for software system design, as stated by Frederick Brooks. Brooks also suggested that Conceptual Integrity can be attained by means of design principles, such as Propriety, and Orthogonality. However, Brooks' principles have not been formalized, posing obstacles to their application in practice, and to a deeper comprehension of Conceptual Integrity. This paper has three goals: first, to achieve deeper comprehension of Conceptual Integrity by deconstructing it into two phases, viz. *Conceptualization* and *Modularization*, iteratively applied during software system design; second, to show that the algebraic Linear Software Models already provide the hitherto lacking formalization of Brooks' design principles, which surprisingly belong mainly to the Modularization phase; third, to reconstruct Conceptualization and Modularization, preserving the desirable tension between: a- *phases' separation*, each with its own specific formal manipulation techniques; b- *precise transition* between these phases, consisting of explicit mutual relationships. The tension stems from the Modularity Matrix linking two very different kinds of entities – system concepts and abstract mathematical constructs – as seldom linked before. The paper motivates the two software design phases, illustrating Conceptualization with examples, and characterizing Modularization by its more mature mathematical theory.

## 1. Introduction

### 1.1 Motivation

Frederick P. Brooks Jr. stated that "*Conceptual Integrity* is the most important consideration in system design", in particular for software systems. In his book "*The Mythical Man-Month, Essays on Software Engineering*" [3], the idea of Conceptual Integrity was first presented. In his subsequent book "*The Design of Design, Essays from a Computer Scientist*" [4], the same idea was re-emphasized and three corresponding design principles were suggested, and verbally explained: viz. Propriety, Orthogonality and Generality. These ideas have been interpreted and applied to software by other researchers.

We have good reasons to agree with Brooks' insight that "*Conceptual Integrity* is the most important consideration for *software* system design". The software word is not meant to be a restriction; it says that in this paper we consider these ideas only within the software

---

Iaakov Exman:   https://orcid.org/0000-0002-9917-3950

context. Indeed, Brooks' account has been motivated by his managing of the development of a large software system at the time, the IBM OS/360 Operating System.

However, the lack of formalization for Conceptual Integrity posed serious obstacles for its systematic application in practice and deeper understanding of the ideas.

### 1.2 The State of the Art

The current situation is a mixture of partially open questions – such as

- What actually is Software Conceptual Integrity?
- What is the role of the design principles?
- How many design principles are needed? Two, three or four?
- How to formalize the whole approach?

and potential answers to some of these questions – such as the growing body of knowledge of Linear Software Models, which is relevant to the interpretation of Brooks' design principles.

### 1.3 This Paper's Goal

The ultimate goal of our work is to obtain a *formal mathematical theory of software* system design for *human understanding* of software systems. It encompasses a theory of software system composition from sub-systems. This theory of software design should be first and foremost applicable in practice to design of any software system of any size. Heuristically, we should somehow combine the above mentioned partially open questions with the potential answers in order to clarify the whole picture. This is done as suggested by the paper title, by conceptual Deconstruction, followed by Reconstruction.

We first deconstruct Conceptual Integrity – in a kind of analysis liberally interpreting the approach of the French philosopher Jacques Derrida [9],[10],[34] – to better understand the very idea. It consists in dissecting Conceptual Integrity into two separate but related phases: software Conceptualization and software Modularization.

Once Conceptual Integrity is deconstructed into two phases, one is able to independently reconstruct each of these phases. We shall see that Brooks' design principles rather belong to the second phase, i.e. software Modularization. Thus, the Linear Software Models are ready to be used as a formalization of the referred design principles. The independently reconstructed phases constitute together our comprehension of Conceptual Integrity.

### 1.4 Paper Organization

The remaining of the paper is organized as follows. In section 2, we present relevant Related Work as preliminaries to the subsequent sections. In section 3, the foundation for our Conceptual Software view is laid down. In section 4, deconstruction of software Conceptual Integrity is performed. In section 5, reconstruction of the $1^{st}$ phase, viz. "Software Conceptualization" is described. In section 6, reconstruction of the $2^{nd}$ phase "Software Modularization" is done. The paper is concluded in section 7 with an overall appraisal and discussion of what has been achieved, and what it is still open to future investigation.

## 2. Related Work

This paper focuses on Brooks' idea of Software Conceptual Integrity and on works derivable in plausible ways from Brooks' idea. As a caveat, the notion of "Conceptual" has been used in additional contexts, quite different from Brooks' ideas. Two examples are given by e. g. Cabot and Teniente [5] who refer to UML/OCL Conceptual Schemas and integrity constraints, and Ganter et al. [22] who describe the "Formal Concept Analysis" domain.

### 2.1 Brooks' Fundamental Idea, Design Principles and Applications

It all starts with Frederick Brooks' idea of **Conceptual Integrity** [3], [4], that originated from his extensive experience with development of large software systems of that time, e.g. the already mentioned IBM OS/360 Operating System. In Brooks' own words: "Conceptual Integrity is the most important consideration in system design" ([3], page 42).

Brooks further details that Conceptual Integrity can be the outcome of a set of three *design principles*: Propriety, Orthogonality and Generality ([4], page 143). These principles were formulated by Brooks in negative terms, as follows: *Propriety*: Do Not introduce what is immaterial; *Orthogonality*: Do Not link what is independent; *Generality*: Do Not restrict what is inherent. These principles have been rephrased in different terms by several authors.

Brooks' design principles were verbally reformulated and illustrated, among others, by D. Jackson and co-authors ([7], page 39): *Propriety* means that a software system should have just the functions essential to its purpose and no more. *Orthogonality* requires that individual functions should be independent of one another. *Generality* demands that a single function should be usable in many ways. Jackson and co-authors illustrated these principles by means of examples (e.g. [28], [29]), in particular a detailed analysis of Git [7], the version control software, producing an improved simpler version which they called Gitless [8].

Despite the lack of formalization, researchers, e.g. Kazman et al., tried direct practical applications with guidance of Conceptual Integrity ideas. Clements, Kazman and Klein in their book [6] refer to Conceptual Integrity as an underlying theme unifying a system design at all its levels. The system architecture should enforce similar things in similar ways, having a small number of data and control mechanisms, and patterns throughout the system. Thus, a more formal definition of Conceptual Integrity would be based upon counting mechanisms and patterns.

Kazman [31] describes a so-called SAAMtool, with visualization capability. *Conceptual Integrity* is estimated by the number of primitive patterns that a system uses. Kazman and Carriere [32] reconstructed given software systems' architecture using *conceptual integrity* as a guideline. Their goal was to attain a restricted number of components, connected in regular ways, with internally consistent functionality. Conceptual Integrity is informally reflected in the restricted number of components, connection regularity and consistent functionality.

## 2.2 Linear Software Models

Linear Software Models have been developed by Exman and collaborators (e.g. [13]) as a formal theory to solve the software system composition problem from sub-systems, down to the simplest architectural units chosen by the software engineer to remain indivisible. Thus, software systems are assumed to consist of a *hierarchy of levels*.

Linear Software Models are based on linear algebra operations and theorems. Each software system level is represented by a Modularity Matrix, whose columns are structors, a generalization of classes, and whose rows are functionals, a generalization of methods or functions. A 1-valued Modularity Matrix element means that its Structor provides the respective Functional.

Making the assumption that all structors are mutually linearly independent and all functionals are also linearly independent – this assumption being motivated by minimization of the number of structors/functionals needed to build the system – a purely linear algebra theorem demands that the Modularity Matrix be square. This is not a trivial result for software systems, as one can easily suggest apparent counter-examples, which are discarded after deeper second thoughts. Indeed it takes some effort to understand the theorem's rationale and implications.

Furthermore, if sub-sets of structors/functionals are disjoint to other sub-sets, a second Modularity Matrix theorem states that these sub-sets can be rearranged into a block-diagonal matrix. These diagonal blocks are recognized as the modules in that software system level (for proofs, examples and further details see the work by Exman [12], [13]).

The modularization of a given design for a software system may not be perfect, having undesirable outliers coupling between modules. One needs a solid theory and a formal procedure to compare different designs of the same software system, and to improve a given design. This is achieved by means of eigenvectors of a suitably weighted and symmetrized Modularity Matrix, as described in [14]. A central theorem for the Modularity Matrix theory is the Perron-Frobenius theorem (e.g. Gantmacher [23]) concerning eigenvectors with all positive elements, fitting the largest eigenvalues of a suitably weighted/symmetrized Matrix.

Exman and Sakhnini [18], [19] have shown how to formally obtain a Laplacian Matrix from the Modularity Matrix. The Laplacian is symmetric by definition and does not need weighting, obtaining the same modules as the Modularity Matrix. One again uses eigenvectors, but with different algebraic theorems. A central theorem for the Laplacian Matrix is the Fiedler theorem [21] for the so-called Fiedler eigenvector fitting the lowest non-zero eigenvalue of the Laplacian Matrix.

Exman and Speicher [20] have also shown the equivalence of the Modularity Matrix to another algebraic structure, the Modularity Lattice, a special case of the Conceptual Lattice which is a basic entity for Formal Concept Analysis [22].

Summarizing, there is a growing body of knowledge on formal techniques to obtain modules for software systems, based upon rigorous algebraic theorems, which are independent of particular software systems, thus independent of software system semantics.

### 2.3 Conceptual Integrity Extension to Agile Design

A recent development of relevance to this work is the perception that agile-design rules, widely known as "the four rules of simple design", first formulated by Kent Beck (see [2] page 57) are similar to the Brooks' design principles. This similarity has been pointed out by Exman [15].

These agile-design rules have been repeatedly reformulated by several authors, with wording and rule-order variations. A slightly rearranged rule-set following Ron Jeffries [30] is: 1- *Test Everything* – All the tests for the SUD (Software Under Development) are passing; 2- *Explicit Intent* – Express the ideas the software's author wants to express; 3- *Eliminate Duplication* – Contain no duplicate code; 4- *Minimize Entities* – Minimize classes and methods.

Among other rule-sets, Corey Haines [26] used the Game of Life to illustrate his rules in a book entitled "Understanding the 4 Rules of Simple Design". Hunt and Thomas [27] mention the design rules in their book "The Pragmatic Programmer", emphasizing in Chapter 2 the relationship between the Duplication rule (their $3^{rd}$ rule) with Orthogonality. In their own words: "The first warns not to duplicate knowledge throughout your systems, the second not to split any one piece of knowledge across multiple system components".

The similarity of agile-design-rules to Brooks' design principles is interesting for two reasons: a- the agile-design-rules are used in practice; b- it links eminently practical approaches to deep foundational considerations.

## 3. Solid Foundation to Formalize Conceptual Software

In this section we lay down a solid foundation, upon which we shall first be able to safely deconstruct a theoretical framework still in its formative stages, and then reconstruct a better understood and more mature design theory of *Conceptual Software*.

### 3.1 Separability of Human Concerns from Software Proper

The idea of "Separation of Concerns" within computing was probably first proposed by Dijkstra in 1974 in his paper "On the role of scientific thought" [11], approximately at about the same time of Brooks' ideas on Conceptual Integrity. "Separation of Concerns" is a *desirable consideration* for software system design, since it is obviously related to modularity. It is in Dijkstra own words "the discovery of which aspects of one's subject matter can be meaningfully studied in isolation for the sake of their own consistency". Dijkstra explains that it involves conscious search – i.e. discovery – of useful and helpful concepts, which is certainly relevant to Conceptual Integrity.

Software Engineering deals with the software system proper as its main subject matter, which gives the name to the engineering discipline. But software engineering also has traditionally dealt with human concerns, viz. its social and economic aspects, such as interactions among teams of developers, and with stakeholders. Indeed, in Brooks' books

Conceptual Integrity of software systems is intermingled with developer teams' considerations.

A relevant Separability Principle [17] is desirable, viz. properties of software proper, in the strict sense, and those of human concerns should be separable. Human concerns should be treated by complementary theories, and not covered by strict design theory of Software proper, the subject of this paper. There are three Separability motivations, all paving the way to neat formalization of Conceptual Software.

First, the *scientific techniques* which are applicable to software proper and to the human concerns have a different nature. Software proper is the subject of a science of the artificial – as discussed in the next sub-section – i.e. the scientific techniques applicable to software are very similar to those of the natural sciences. On the other hand, human concerns are usually dealt with by techniques of the social sciences, such as economics and sociology. Citing Dijkstra again, "focusing one's attention upon some aspect does not mean ignoring other aspects". Separability is for the benefit of all separated aspects.

Second, *formal verification* of correctness is a rational requirement for software system design. Verification should be independent of which team of engineers developed the software system or which stakeholders are the customers of the referred system. Suppose two identical copies, of a software system, are requested by safety net considerations, to be produced by two different industrial manufacturers, with two different teams. It should be clearly possible to verify the correctness of each copy by the same procedure, independently of the developer teams. Moreover, the verification of each copy may possibly give different results from the other one. Separable verification is the common assumption for products of all engineering disciplines, other than software engineering.

Third, *autonomous software* (and embedded) systems development by machine learning and robotics, with self-maintenance and self-further-development capabilities are increasingly common. For instance, a satellite navigating to a remote planet should be able to take immediate control decisions, self-check and upgrade itself, independently of the attention of any human being at a remote distance, too impractical for real-time cooperation. The same could be said about dangerous environments on planet Earth. We envisage Separability of theories as a pre-condition for completely autonomous systems, for which formal verification is an even more stringent demand.

### 3.2 Software: A Science of the Artificial

The somewhat paradoxical nature of the formal techniques relevant to software is that of a science of the artificial. On the one hand, software is a creation of humans, expected to behave as it was planned. Indeed, novice programmers have absolute confidence on the correctness of the simple program that was just written, compiled and run.

On the other hand, what a surprise! The program either fails miserably or produces undecipherable results. This is a recurring observation, accumulated by the experienced programmer. It takes experience with software to understand that it is rather complex and has obvious similarities to the subjects of natural science, as if it were not designed by humans.

For large software systems, the situation is only aggravated. Software has its own laws and should be tested and measured in order to discover the contents of these laws.

We refer the reader to the nice book by Herbert Simon [36] concerning the scientific method legitimacy for "The Sciences of the Artificial". As an example ([36], page 7) an airplane – an artifact planned and produced by humans according to aeronautical engineering techniques – and a bird – an animal found in nature which is able to fly due to its wings – can both be analyzed by natural science methods, since both obey the laws of aerodynamics.

### 3.3 A Mathematical Theory: Why Algebra?

In our vision, we need a *rigorous mathematical theory* behind the whole Conceptual Integrity approach to software system design. No theory should be totally abstract and/or devoid of intended application in practice. We consider previous experience and applicability as means for validation of the theory proposed in this work.

Why mathematics? The main justification is to have a uniform and generic approach to any software system whatsoever. Any, means any size, any kind of application, any kind of hardware system in which the software may be embedded.

We shall use existing mathematical domains and theorems, eventually in novel surprising contexts. This is the usual scientific method which is now applied to software. This approach is in complete analogy to standard mathematics e.g. behind Maxwell equations of electro-magnetism. The real challenge is to work out the formalism to check whether the theory indeed predicts the accepted wisdom of software engineering.

Why algebra? There exist a variety of kinds of algebraic structures, and these structures are flexible enough, to allow manipulation of the seemingly inexhaustible variability of software systems and sub-systems. The literature shows the relevance and often equivalence of matrices, lattices, graphs and other structures. Last, but not least, algebra seems adequate to treat together the two kinds of entities relevant to software design – software system concepts and abstract mathematical constructs – as discussed next.

### 3.4 Software System Concepts and Abstract Mathematical Constructs

We conclude this section posing the deep problematic of Software Conceptual Design that probably has been the main impediment to development of an actual theory for Software Design. The problematic is the intimate interaction between two very different kinds of entities:

- *Software System Concepts* – naturally appearing when one defines and characterizes a software system;
- *Abstract Mathematical Constructs* – such as mathematical graphs, vectors and various algebraic structures, without which there is no possible formalization – i.e. no theory – of software system design.

Our claim is that the very problematic of software design is also the source of its solution!

The intimate interaction between the above two kinds of entities, on the one hand should explicitly appear in the software system representation; on the other hand the two kinds should not mutually interfere in the formal manipulation procedures which are specific for each kind of entities.

The Modularity Matrix is an example of a structure that embodies that interesting and desirable characteristic of our foundation for a theory of software design, worth of being explicitly highlighted viz. the Matrix contains the two very different kinds of entities, as seldom linked before.

One kind of entities is *a pair of sets of concepts* – which are naming the structors and functionals – consisting in the highest abstraction description of a specific software system. For instance, typical concepts for a banking ATM (Automatic Teller Machine) are *structors* like bank-account, touch-screen, security-unit, and respective *functionals* like withdraw/deposit-cash, touch-to-choose-operation, encrypt/decrypt-message. These two sets of concepts justify coining the software system as *Conceptual Software*.

The other kind of entities is the *sets of numerical column and row vectors* in the Matrix, corresponding to the above structors and functionals, which display the explicit relationships between them. The numerical vectors enable *strict application of linear algebra* operations to formally obtain software modules. These relationships, although numerically represented, are also exactly those needed to convert *terms* into actual *concepts*. So, the two different kinds of entities – illustrated in Fig. 1 – are indeed intimately related.

| Structors → <br> Functionals ↓ | | Bank-account <br> S1 | Checking-Account <br> S2 | Savings-Account <br> S3 | Touch-Screen <br> S4 | Security-Unit <br> S5 |
|---|---|---|---|---|---|---|
| Open-Account | F1 | 1 | 1 | 1 | | |
| Withdraw/Deposit-Cash | F2 | 0 | 1 | 0 | | |
| Calculate-Interest | F3 | 0 | 0 | 1 | | |
| Touch-to-Choose-Operation | F4 | | | | 1 | |
| Encrypt/Decrypt-Message | F5 | | | | | 1 |

**Figure 1** – A simplified ATM Modularity Matrix – It has 5 structors (columns) S1 to S5, and 5 functionals (rows) F1 to F5. A 1-valued matrix element means that a structor provides the respective functional. For instance, Structor S3 (Savings Account) provides two functionals F1 (Open-Account) and F3 (Calculate-Interest). There are three modules (in blue background) in this matrix: upper-left is the Bank Accounts 3*3 module; middle is the Touch-Screen 1*1 module; lower-right is the Security-Unit 1*1 module. In this matrix all elements outside the modules (in yellow background) are 0-valued, but the outside zeros are omitted for clarity. The Bank Accounts module has three 1-valued matrix-elements in the same F1 row, implying inheritance of the Open-Account functional. The S1 Bank-account is a generic structor (i.e. an abstract class in programmers parlance), while Checking-Account and Savings-Account are more specific ones (i.e. sub-classes of the abstract one). Note: all figures online are in color.

These two kinds of entities justify, on the way to a formalized Conceptual Software, the Deconstruction of Brooks' Conceptual Integrity whole idea into two separate ideas. This is, in simple terms, motivated by the need to:

- *disentangle to a certain extent* the intimate interactions between the above kinds of entities, but not totally eliminate them;
- *enable specific formal manipulations* for each kind of entities, be they conceptual or algebraic.

Deconstruction is done in the next section.

## 4. Deconstruction: Conceptual Integrity of Software

Liberally following Jacques Derrida's [9],[10] approach, we first deconstruct Conceptual Integrity, to have a clearer and deeper understanding of Brooks' idea. In our viewpoint deconstruction is only desirable if followed by a reconstruction effort.

### 4.1 Two-phase Software Design

The single idea of "Conceptual Integrity" is here deconstructed into two ideas, viz. *Conceptualization* and *Modularization*. These ideas are immediately used as two phases of a software system design procedure. These are shown in a pseudo-code format in the following:

> *Software Design* **Procedure:** from **Conceptual Integrity**
>
> **{Set** *Modularity Criterion*;
> **Repeat until** (software system modules obey *Modularity Criterion*)
>    **{   1$^{st}$  phase:** Software System **Conceptualization;**
>    **2$^{nd}$ phase:** Software System **Modularization;   }   }**

From now on we shall use *Software Design Procedure* as the procedure name, omitting the word system, but always having in mind that software implies a *Software System*.

*Conceptualization* is a choice of concepts essential to describe a software system. One starts with an initial proposal and may add/delete concepts until the procedure's termination, when it reaches the Modularity Criterion. Termination must occur, even in cases of frequent conceptualization changes. We do not assume continuous design. *Modularization* performs clustering of sub-sets of concepts into software system independent modules, which are reasonable from a Conceptualization viewpoint. Modularization helps to focus conceptualization. Figure 2 illustrates one cycle of Software Design, using the ATM example of Fig. 1.

In the 1$^{st}$ *Conceptualization* phase, the software engineer:

- ***Proposes concepts*** – the two sets of concepts, serving as Structors and Functionals;

- ***Determine concepts' relationships*** – i.e. which structors provide their respective functionals that constitute the numerical matrix.

In the 2<sup>nd</sup> *Modularization* phase, the software engineer:

- ***Uses the matrix without concepts*** – the numerical matrix is the input to the modularization phase, temporarily ignoring the concepts;
- ***Produces best modules*** – these are the output of the modularization phase, which then are recombined with the two sets of concepts.

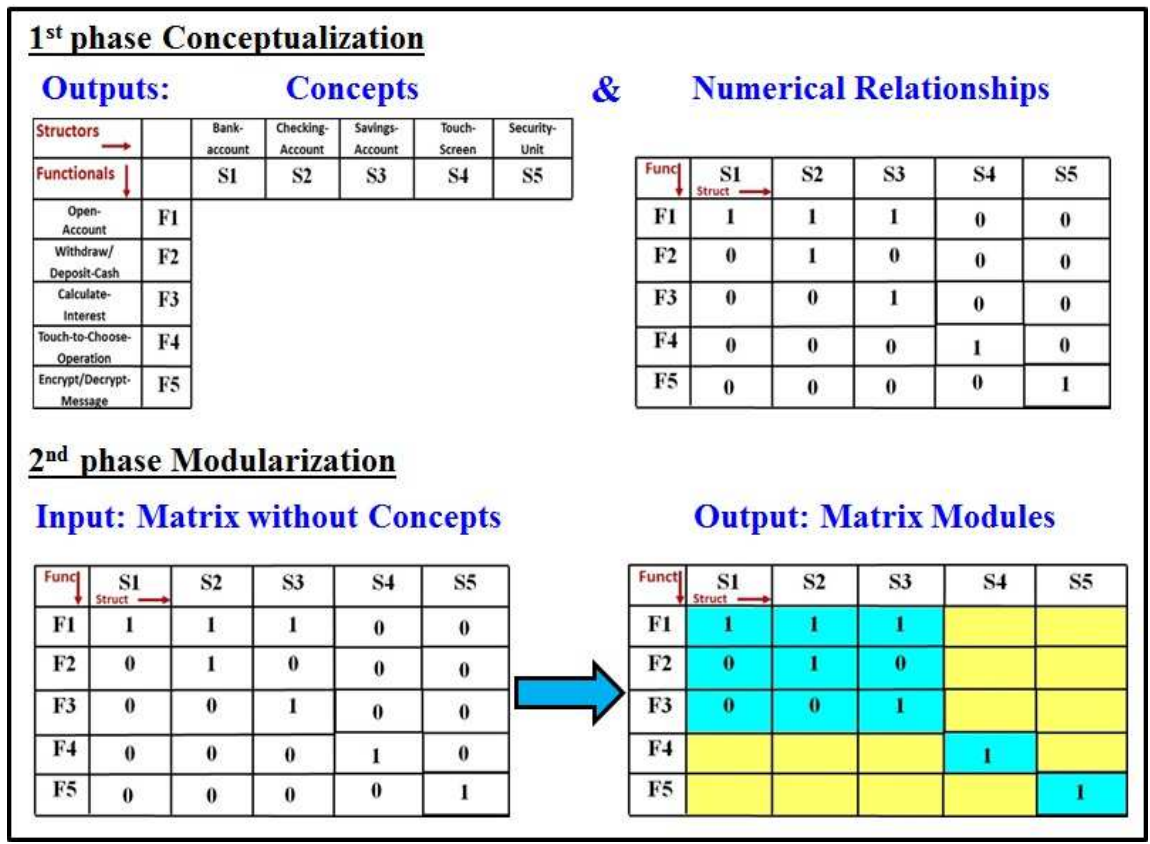

**Figure 2 –** The *Software Design* **Procedure** illustrated by a Simplified ATM Modularity Matrix – The upper part of this figure shows the 1<sup>st</sup> *Conceptualization* phase outputs. A designer proposes Structor and Functional concepts, and determines which functionals are provided by their respective structors. The lower part of this figure shows input/output of the 2<sup>nd</sup> *Modularization* phase. The designer inputs the Relationships matrix from the previous phase, temporarily ignoring the concepts. This phase output is the set of modules of the software system being designed. After the 2<sup>nd</sup> phase, one reassembles the concepts with the modularized system into the full Modularity Matrix.

### 4.2 Analysis of Conceptual Principles of Integrity

The Deconstruction process of Conceptual Integrity is not concluded by the partition of the Software Design Procedure into two phases, as done in the previous sub-section. One still needs to analyze the design principles proposed by Brooks.

We first look at the ***Propriety*** design principle in sub-section 2.1 (3$^{rd}$ pagragraph). Substituting the "functions" by "concepts" to deal with it in most general terms, one obtains: "*Propriety* means that a software system should have just the concepts *essential* to its purpose *and no more*". One perceives that this principle conveys two distinct kinds of meaning:

- ***Essentiality of the chosen concepts*** – as stated explicitly through the word "*essential*";
- ***Restriction of number of concepts*** – as expressed in "just the concepts needed to its purpose *and no more*";

We claim that essentiality of concepts is not related to their number. Different designs of the same software system may have the same number of differing concepts, some of them being essential, while other ones being incidental. <u>Essentiality of concepts</u> is naturally related to the 1$^{st}$ phase of the Software Design Procedure, viz. ***Conceptualization***. An essential concept is indispensable for a given software system. Moreover, it should have a quality yet undefined – i.e. not found in the Propriety design principle – of having mutual Conceptual Integrity with other concepts, i.e. a positive quality of being essential.

On the other hand, the <u>restriction of number of concepts</u> naturally belongs to the 2$^{nd}$ phase of the Software Design Procedure, viz. ***Modularization***, which is a general and exact process to simplify and reduce the number of concepts and cluster them into separable modules.

Next, we look at the ***Orthogonality*** design principle (sub-section 2.1, 3$^{rd}$ paragraph). Again, replacing "functions" by "concepts": "*Orthogonality* requires that *individual concepts* should be *independent* of *one another*". The words "*of one another*" seem to be redundant, deserving analysis. This principle has two distinct kinds of meaning needing deconstruction:

- ***Individual Concepts should be independent*** – individual concepts at their specific abstraction level should be independent;
- ***Concepts independent of one another*** – if all individual concepts were mutually independent to the same degree, there could not be modularization, i.e. clustering of related concepts into modules. Concepts of a module relative to concepts of all other modules are <u>independent of one another</u> to a higher degree, due to their mutual orthogonality.

## 5. Reconstruction 1$^{st}$ Phase: Software Conceptualization

In this section we begin to perform Reconstruction of the 1$^{st}$ phase of the Software Design Procedure. For a certain software system the overall idea of Software Conceptualization is:

- *Choice of essential Concepts* – an initial definition of the software system;
- *Choice of relevant Domain Ontologies* – based upon the chosen essential concepts, aiming to generate the software system *Application ontology*;
- *Assignment of Relationships among Structors and Functionals* – needed for the Modularization phase.

The next sub-sections describe assumptions regarding essential concepts, domain and application ontologies. These are followed by definitions necessary for Conceptualization. The elementary conceptual kinds (undefined, but explained) in this section are: domains, ontologies, architectural concepts, attributes and their ranges of values.

### 5.1 Software Conceptualization Assumptions

In order to reconstruct Software Conceptualization, we make some basic assumptions:

- *Preliminary Understanding by the Software Engineer* – the software engineer responsible for the design should have an intuitive *preliminary* understanding of the software system to be designed and later developed: one is able to distinguish concepts that belong to the software system from those that are beyond the scope of the system. This understanding is needed to start designing. Conceptualization and Modularization iterations increase understanding of the software system.
- *Computation for Understanding* – despite the deconstruction of Conceptual Integrity into two distinct phases, these have in common the ubiquitous theme of "small numbers". The rationale for small numbers is enabling efficient computation of needed quantities, for *deep understanding* of the software system by human stakeholders.
- *Concepts found in relevant Domain Ontologies* – preliminary understanding also enables to determine the relevant Domain Ontologies (see e.g. Guarino [24] for the computational, not the philosophical notion of ontology). For instance, for the ATM system in Fig. 1, these Domain Ontologies can be chosen as: Financial domain (for bank accounts); Human-machine-interface domain (for the touch screen); Communication-&-security domain (for the security unit).

    Conceptualization from widely adopted Domain Ontologies is justified by:
    a- *An accepted Common Vocabulary* – to avoid arbitrary choices of terms, not recognized by other practitioners in the same domain;
    b- *Self-consistency is not enough* – beyond arbitrary individual term choices, a more serious problem is a whole vocabulary which is overall self-consistent, i.e. still displaying conceptual integrity, but not conforming to common usage meanings of terms. This is a severe hindrance to scientific exchange, as observed by Plebe and Grasso [35]. For instance, imagine a random permutation among terms of a Human-machine-interface domain, within a given software system. Suppose that a *window* has exchanged meanings with a *computer screen*, i.e. a window would mean a fixed size computer glass screen and a screen would mean a resizable and movable window. Indeed self-consistency is not enough.
- *Application Ontology as specialization of Domain Ontologies* – an intermediate outcome of the Conceptualization for the software system under design is an Application Ontology, i.e. a specialization of the relevant Domain Ontologies. An Application Ontology (e.g. Guarino [24] Fig. 4, pages 7-8) describes concepts related to certain Domains and specific Tasks. For the ATM example, a "checking account" is a concept of both the Financial Domain Ontology and the

- specialized ATM system Application Ontology, while a "mortgage" is a concept of the Financial Domain Ontology, but not of the ATM Application Ontology.
- ***Concept Changes at the lowest possible abstraction level*** – concept changes along a software system history are a matter of fact. Therefore, one should choose higher abstraction level concepts with extreme care, to push eventual concept changes to lower abstraction levels, to the maximal possible extent.
- ***Adopt the FOREST view of formal Conceptualization*** – to avoid the syndrome of "can't see the forest for the trees", we prefer deep understanding – to see the Forest – instead of detailed formal Conceptualization. We avoid adopting for now a specific ontology language: neither OWL, (or its sub-species OWL-lite, OWL-DL, OWL-FULL) nor RDF, RDFS, or the Protégé tool terminology; their syntactic burden, the time wasted in details, and eventual controversies, obscure the understanding of Conceptual Integrity. This does not preclude a few definitions in the next sub-sections and theorems elsewhere, minimizing details. One should be able to translate, later on, our definitions into a suitable ontology.
- ***Strictly formal TRANSITION between the two phases*** – despite the Forest view of formal Conceptualization, there is a *strictly formal transition* between the Conceptualization phase and the Modularization phase. The transition enables exact algebraic methods within the Modularization phase. The deconstruction into two phases is largely motivated by the possibility of the strictly formal transition, between the two phases.

**5.2 Highest Essential Concepts: a Software System Abstraction**

Here we finally try to cope with the fundamental question:

- What actually is the source of *intrinsic* Conceptual Integrity?

We aim to define Essential Concepts, which appeared in the formulation of the Propriety design principle. We first motivate this notion informally. Any Essential Concept must be present in the Software system Conceptualization, thereby sharpening the system boundary; otherwise the software system would lack integrity relatively to the concepts already included in the system.

Now, the surprising statement: there is *nothing intrinsically essential in a specific concept* for Conceptual Integrity in the *highest abstraction level* of any software system. Human engineers can imagine and design any possible kind of system, from very strange ideas, less successful products, to widely adopted successful systems. And in between one finds many intermediate options.

For example, there have been a huge variety of *airplanes* of various sizes, engine numbers, passenger numbers, number of decks, distance ranges, etc. But, one can find also much less widespread *seaplanes* – that take off and alight on water – and *amphibious aircraft* that may take-off and alight on both land and water. Does a strict airplane have more Conceptual Integrity than an amphibious aircraft? These systems have clearly differing functionalities, but amphibious aircraft have been imagined, designed, manufactured and used without problems. A theory of conceptualization must take into account all the possible varieties of working systems.

We define Essential Concepts assuming that software systems are hierarchical (see sub-section 2.2 on Linear Software Models). This allows us to separate two kinds of essentiality: one for the highest hierarchy abstraction level and another one for all lower subsequent levels, i.e. whose essentiality follows from conceptual integrity.

> **Definition 1: Essential Concepts of the Highest Abstraction Level**
>
> The Essential Concepts of the highest abstraction level of an Application Ontology of a given Software System are defined by two characteristics:
> - a- Each Essential Concept is *arbitrarily chosen* by the human software engineers responsible for the overall system design;
> - b- There are *small numbers* of these essential highest concepts, of the order of a few units.

The Highest Essential Concepts, that populate the highest abstraction level of an Application Ontology, are an abstraction of the given Software System. The Application ontology of this software system is gradually built from the Domain Ontologies relevant to the Highest Essential Concepts. The highest abstraction level of the Application Ontology does not necessarily correspond to the highest level of each source Domain Ontology.

### 5.3 Conceptual Integrity within the Application Ontology

The Essential Concepts in all subsequent lower abstraction levels of a software system and its corresponding Application Ontology levels are not anymore arbitrary. They are coined Essential *Integrity* Concepts as they follow by Conceptual Integrity from the class hierarchy within each source Domain Ontology. For instance, for the ATM system the highest essential concept within the Financial Domain is e.g. arbitrarily chosen as "bank-account". Then, the next levels, i.e. sub-classes of the highest essential concept, are sub-classes within the Financial Domain ontology, e.g. "savings-account". Here is the generic definition:

> **Definition 2: Essential *Integrity* Concepts of a Software System**
>
> The Essential Concepts of the all subsequent lower abstraction levels, except the highest one, of an Application Ontology of a given Software System are defined by two characteristics:
> - a- Each Essential Concept follows by *Conceptual Integrity from the sub-classing* hierarchy of the source Domain Ontology;
> - b- There are *small numbers* of essential integrity concepts in each abstraction level of the Application Ontology hierarchy, of the order of a few units.

As one perceives:

- *Conceptual Integrity* of a Software System is heavily dependent on the rational preparation of each Domain Ontology;
- *Small numbers* is a basic demand systematically appearing in all our definitions.

### 5.4 Concept Characterization within the Application Ontology

There are two motivations to characterize concepts within the Application ontology of a software system. First, is to guarantee distinguishable concepts. There should not be different ontology terms standing for identical concepts (unless declared as synonyms, a finesse not concerning us at this formalization stage). Concepts are distinguished by their attributes, and ranges of values admissible for each attribute. Second, attributes appear in modelling languages, such as UML, and in object-oriented programming languages used to implement software systems. Next we define concept characterization.

---

**Definition 3: Concept Characterization in a system Application Ontology**

A Concept Characterization is an ordered pair whose first element is the name of a "*domain*" and whose second element is a "*tuple*", itself consisting of ordered pairs, each containing the name of an "*attribute*" and a legal "*range of values*" for the respective attribute, obeying the following rules:

a- Its "*domain*" belongs to one or more Domain Ontologies used to generate the software system Application Ontology;
b- It contains the minimal *small numbers* "*tuple*" needed to distinguish it from all other concept characterizations in the same Application ontology.

A Concept Characterization is represented as:

Characterization(*concept*) =
  {*domain*, ([*attrib$_1$, range$_1$*], [*attrib$_2$, range$_2$*],…,[*attrib$_k$, range$_k$*])}

---

*Domain* is the name of a source Domain Ontology. Minimal "*tuple*" size means the minimal number of attributes and their respective ranges of values. The "*range of values*" can be a type of numbers (e.g. integer, real, etc.), a specific number, a set of numerical intervals, a set of literals. This definition has some desirable computational properties:

- *Domain assignment* – it is easily computed, while each concept of the Application Ontology is generated from the respective Domain Ontologies;
- *Distinguishability of Concept Characterizations* – it is computed by comparing the respective *tuples'* attribute names and their range of values; the minimal *tuple* size for overall distinguishability within a given Application Ontology is also efficiently computed.

Concept Characterizations are not unique, as we do not propose fixed rules for assigning the number of attributes or their ranges of values. Concept Characterizations may change in time, in response to introduction of new concepts in the Application Ontology. For instance, when dealing with the domain of vehicles, possible concept characterizations are:

Characterization(*bicycle*) = {*vehicle*, ([*wheels, 2*], [*tire-width, "narrow"*], [*engine, "none"*]}

Characterization(*motorcycle*) = {*vehicle*, ([*wheels, 2*], [*tire-width, "wide"*], [*engine, "one"*]}

In these examples, one could put a range of numerical values for the "tire-width" attribute instead of literals. One also could put numerical values for the "engine" attribute, or alternatively describe its power, e.g. in terms of "horse-power".

Superfluous concepts cannot be distinguished from other concept characterizations, i.e. one has different terms for the same concept, in the same Application Ontology. Superfluous concepts should be eliminated from the Application Ontology of a software system. Incidental concepts are neither essential, nor superfluous; they will be dealt with elsewhere.

### 5.5 Summary of Conceptualization Steps

In order to perform Conceptualization, one needs a series of more detailed actions (see e.g. Exman and Iskusnov [16]). These are summarized as follows:

a- ***Choice of Essential Highest Concepts*** – which represent a starting definition of the software system;
b- ***Choice of Domain Ontologies*** – decide which Domain Ontologies are relevant to the system being designed;
c- ***Resolution of Ambiguities*** – resolve any ambiguous term which may appear in different domains. For instance, *liquidity* may refer to financial assets readily converted into cash, or to the physical state of the matter, e.g. water in room temperature is a liquid;
d- ***Choice of Essential Integrity Concepts*** – choice of sub-classes of the Essential Highest Concepts, within the respective Domain Ontologies, given suitable criteria;
e- ***Generation of the Application ontology*** – obtain the application ontology, from the chosen Domain Ontologies, and respective Essential Integrity Concepts.
f- ***Initial Assignment of Structors and Functionals Relationships*** – this assignment builds the purely numerical part of the Modularity Matrix, as input to the $2^{nd}$ phase of the Software Design Procedure. This initial assignment should be refined by the $2^{nd}$ Modularity phase.

In order to achieve a formal theory of the Conceptualization phase, one must apply techniques of a suitable branch of mathematics, such as a formal algebra. A formal algebra could be obtained e.g. from the translation of the Application Ontology, to a chosen formal ontology language. Such a translation is better done by automating it with an especially built software tool, as one may need to modify existing algebras to some extent. A comprehensive formal theory of the Conceptualization phase is beyond the scope of this paper.

### 5.6 Conceptualization Change Examples

We discuss here software systems illustrating different situations with respect to essential concepts and changes along history. ATM is a currently widely used system, displaying a software system definition variable along time. A Clock is a system with a very long history and very stable definition, and myriads of specific implementations. The Airline Flight Reservation system illustrates a system conceptualization with conflicting interests among stakeholders, causing frequent conceptualization changes.

### 5.6.1 ATM Software System

An informal definition of an ATM is a system in a static location that enables operations on a remote bank-account by means of a human-machine-interface linked to the bank-account by a communication-and-security network. The three Essential Highest Concepts are those mentioned in this informal definition, and corresponding to the modules in Fig. 1: *bank-account*, *human-machine-interface*, and *communication-and-security*.

Since the *bank-account* module has been slightly expanded in Fig. 1, one finds two next level Essential Integrity Concepts: *checking-account* and *savings-account*. These concepts are a simplified example of a much more complex system. A concise sample of references to the corresponding domain ontologies are: Banking-and-Finance ontology [1] including all the concepts mentioned in this example, viz. ATM, bank-account, checking-account, savings-account, mortgage; and Security ontologies [25], see also references therein.

ATM software systems have changed along their existence, to keep systems up-to-date. For instance, earlier ATMs allowed only local currency operations. More recent ones also allow foreign currency operations. System updates illustrate the need for "Deconstruction, then Reconstruction" of concepts. It also shows that a good choice of the highest level essential concepts, avoids frequent changes at that level. The abstract *bank-account* concept encompasses both local-currency-accounts and foreign-currency-accounts.

An ATM system is not a customary way to obtain a mortgage to buy a house, i.e. "authorize-mortgage" is not an allowed operation through an ATM. An explanation is given by the relevant concept characterizations (by Definition 3 above) of an allowed operation such as "cash-withdrawal" and a forbidden operation such as "authorize-mortgage":

Characterization(*cash-withdrawal*) = { *bank-operation*, ([*type, cash*], [*amount, "cash-limited"*], [*duration, "one-day"*]}

Characterization(*authorize-mortgage*) = {*bank-operation*, ([*type, loan*], [*amount, "collateral-property-limited"*], [*duration, "long-term"*]}

One sees the two significant differences: the amount and duration properties of the "authorize-mortgage" do not fit the nature of the current ATM usage.

Another question of interest is the arbitrariness of implementations. Is a touch-screen or a push-button essential for the ATM? One can have a system with just a push-button interface, or only touch-screen interface, or a combination of both. The latter one is justified by touch-screen ease-of-access for the wide public and the push-button as a consideration for elder people used to previously used interfaces. None of these options is essential, illustrating the arbitrariness referred to in Definition 1.

### 5.6.2 Clock Software System

A clock is a very old system, whose definition has stabilized along the humankind history. A modern informal definition of a clock is a device to display time in a given numerical scale, based upon a periodic physical phenomenon, allowing it to be synchronized with other clocks. One has three Essential Highest Concepts just mentioned in this informal definition: *numerical-scale-display*, *periodic-phenomenon*, and *adjust-for-synchronization*.

Old stable systems, with a long history definition, are almost not arbitrary anymore. The *scale-display* is known from sundials since the Egyptian and Babylonian astronomies. The *periodic-phenomenon* is known from the same ages, as sundials reflect the apparent periodic motion from sunrise to sunset. The third Essential Highest Concept, the *adjust-for-synchronization* is somewhat newer, probably from the middle-ages' cathedrals' clock-tower.

The huge variation of implementation technologies, sizes, precision, etc. are obviously abstracted in the highest Application Ontology concepts. These can be derived from the Domain ontology source, which in this case is e.g. a standard Time Ontology [37]. An example in this ontology of a Time next level Essential Integrity Concept (sub-class) is *Time-Zone* which is needed for *adjust-for-synchronization*.

Despite the long conceptual stability, a recent conceptual change is the definition of the *second* time unit. The old historical concept is a solar day division into 24 hours, each hour of 60 minutes and each minute of 60 seconds. Recent concepts, with much higher precision serving as a globally available standard, are defined in terms of other periodic phenomena e.g. of a Cesium atomic clock. The new concept, with a seemingly arbitrary time conversion constant, preserves the duration of the older concept of a second.

### 5.6.3 Airline flight reservation Software System

The Airline Flight Reservation system conceptual model is an example pointing out to a conceptualization problem caused by conflicting situations among software system stakeholders. An informal definition of the Airline Flight Reservation is a system that enables advance reservation and later management by a passenger, of a journey involving requested airlines, flights and airports. Thus, one has here five Highest Essential Concepts: 1-passenger; 2-journey; 3-airlines; 4-flights; 5-airports (cf. e.g. [28] page 7). The journey consists of a series of flights. The flights involve one or more airlines, among a few airports.

This choice of highest essential concepts is a simplified system. For instance, we left out a travel agent that may provide various other services, such as Hotels, ground transportation, and car rental, significantly increasing the complexity of the Airline Flight Reservation. Such choices are not unique: one may compare the above concepts with those found in two different ontologies for aviation by Keller [33] and by Vukmirovic et al. [38]. Examples for the *airlines* concept (class), from the ontology referred to in [38], of possible next level Essential Integrity Concepts (sub-classes) are *Marketing-Airline* and *Code-Share*.

System Conceptualization with conflicts among stakeholders implies further complexities. In this flight reservation system one finds an airline vs. passenger conflict, i.e. the airline wishes to maximize profits, while the passenger wishes to minimize prices. A way to maximize profits is a system of flight classes allowing the airline to sell, to different passengers in the same flight, tickets at different prices, of which passengers are often not aware. This causes frequent conceptualization changes, often not easy to follow.

## 6. Reconstruction 2$^{nd}$ Phase: Software Modularization

In this section we perform Reconstruction of the 2$^{nd}$ Phase of the Software Design Procedure, viz. the Software Modularization phase. The overall Reconstruction idea consists in fully adopting the Linear Software Models, which were independently developed, as the algebraic theory of Modularity corresponding to Brooks' ideas of Conceptual Integrity. The justification consists in the following steps:

- *Reformulation of Brooks' principles into Algebraic Principles of Integrity* – this reformulation is done, assuming the plausibility of the Deconstruction analysis of the conceptual principles of integrity (in section 4.2);
- *Adoption of Matrix models of software systems* – since the Linear Software Models' representations of software systems are generated by application of the above referred algebraic principles of integrity to the relevant matrix;
- *Modularization by standard linear algebra on the matrices* – for instance, spectral methods obtaining eigenvectors, which are a suitable formal technique taking the matrix models as representations of the software systems.

This Reconstruction obtains an Algebraic Software Modularity Theory which is more mature than that of the Conceptualization phase. This theory has been implemented upon either the Modularity Matrix or the Laplacian Matrix.

The input to the Software Modularization Phase is the purely numerical Modularity Matrix obtained in the final step "*f*" of the Conceptualization Phase – see sub-section 5.5 and Figure 2. The Modularization Phase itself is semantically blind. At the end of this 2$^{nd}$ Phase, the full matrix is reassembled with the structor and functional concepts, and semantics may serve to verify the attained Modularity. Another design cycle is performed, until the stopping *modularity criterion* is achieved, as seen in the Software Design Procedure in sub-section 4.1.

### 6.1 Brooks' Principles Reformulated: Algebraic Principles of Integrity

Application of the Brooks' propriety and orthogonality design principles for Conceptual Integrity as algebraic constraints to the numerical Modularity Matrix, at a given level of a software system, obtains two basic theorems of the algebraic theory [13], as stated in the Related Work review of Linear Software Models in sub-section 2.2. These theorems are applied in the 2$^{nd}$ Modularity Phase of the Software Design Procedure. *Generality*, the 3$^{rd}$ Brooks' principle, seems not to be an independent design principle, and will be dealt with elsewhere.

Brooks' *Propriety*, which is supposed to reduce the number of superfluous concepts, is reformulated as the **algebraic demand of Propriety**, producing the desired reduction effect by **linear independence**. The explicit algebraic demand is that all the Matrix structors, the column vectors, be linearly independent, and all the Matrix functionals, the row vectors, also be linearly independent. Modularity Matrix' propriety is thus formally measured by the matrix rank. The algebraic demand of *Propriety* implies that the Modularity Matrix should be a square matrix. This is a pure linear algebra theorem, which is true only if the hypotheses, linear independence of both structors and functionals, are fulfilled. As already mentioned in sub-section 2.2, it takes some time to understand the software requirements and implications of this theorem.

Brooks' *Orthogonality* is reformulated as the **algebraic demand of Orthogonality** among modules. This algebraic demand implies that if sets of structors and corresponding functionals are disjoint to other sets of structors and their functionals within the Modularity Matrix, they can be reordered as a block-diagonal matrix. Diagonal blocks are recognized as the software modules, since structors and corresponding functionals in a given module are respectively *orthogonal* to structors and their functionals in all other modules at that system level. The colloquial usage of *orthogonality*, such as in Brooks' design principles, imply "sharply divergent"; it comes from mathematics, meaning at right angles. Orthogonality of a pair of vectors is measured by their scalar product. This measure should be extended to all pairs of row vectors and pairs of column vectors, in order to measure the whole Matrix orthogonality. Orthogonality is a stronger requirement than linear independence, since linear independent vector sets are not necessarily orthogonal, while orthogonal vectors are necessarily linear independent.

Block-diagonality is a source of formal design criteria that we were looking for. Outliers, i.e. 1-valued Modularity Matrix elements outside modules, coupling the respective modules, imply a lack of Modularity, and also point out to problematic spots demanding software system redesign. The theory provides intrinsic measures of design quality, and also guides the software engineer towards design improvement.

### 6.2 A Unified Algebraic Theory of Software Modularity

The second step of the Reconstruction of Software Modularization is the adoption of algebraic – in particular matrices – models of software system. The algebraic theory provides rigorous formal techniques to obtain module sizes for any given software system. These "*spectral*" techniques start from a numerical matrix, such as that obtained from the output of the 1$^{st}$ Conceptualization Phase of the Software Design Procedure. Then one calculates the matrix eigenvectors. The Modularity Matrix must be symmetrized and suitably weighted before calculating eigenvectors. Modules are obtained as non-sparse connected components, as schematically illustrated in Fig. 3.

Similar spectral techniques are applicable to the Laplacian Matrix [18], [19] obtained from the Modularity Matrix through a structor-functional bipartite graph. The Laplacian is symmetric by construction and does not need weighting. The Modularity Matrix is also equivalent to another algebraic structure, the Modularity Lattice [20]. Finally one has at one's hands a unified algebraic software theory of modularity, of which all these structures are representations.

### 6.3 Modules from the Modularity Matrix

Using the Modularity Matrix, modules are revealed by the eigenvectors [14] of the symmetrized and weighted matrix, as derived from the Perron-Frobenius theorem [23]. A schematic illustration of the Modularity Matrix is seen in Fig. 3.

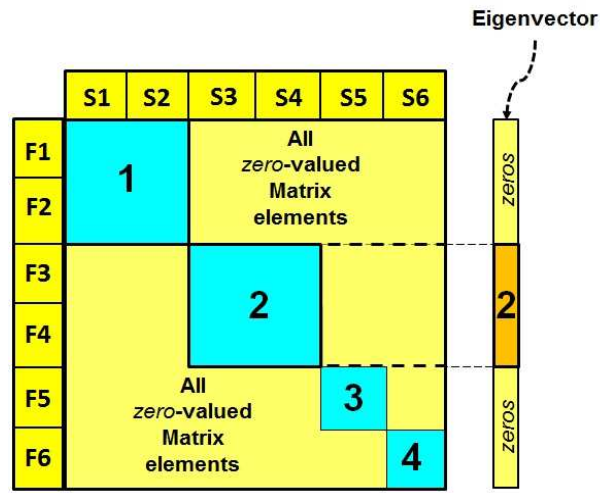

**Figure 3**. – Schematic block-diagonal Modularity Matrix with eigenvector – This matrix has six structors (S1 to S6), six functionals (F1 to F6) and four numbered modules (blue background). Most of elements inside the modules are 1-valued and some of them are zero-valued. All matrix elements outside the modules are zero-valued as there are no outliers coupling between modules. Eigenvectors calculated from the symmetrized and weighted Modularity Matrix determine the size of its modules, one eigenvector per module. This is illustrated for module #2 by its non-zero eigenvector elements (in orange).

### 6.4 Modules from the Laplacian Matrix

An alternative matrix model of software systems is the Laplacian Matrix which is derivable [19] in two steps from the Modularity Matrix:

a- ***Generate a Bipartite Graph from the Modularity Matrix*** – a Bipartite Graph links a set of structors (S1 to S6 in Fig. 3) with a set of functionals (F1 to F6 in Fig. 3), such that vertices in one set are only linked to vertices in the other set; each Matrix element which is 1-valued obtains one Bipartite Graph edge, i.e. from one structor to one functional.

b- ***Get the Laplacian Matrix from the Bipartite Graph*** – this is done by the formula:

$$L = D - A \qquad (1)$$

where ***L*** is the Laplacian matrix, ***D*** is the Degree matrix, a diagonal matrix showing the degree of each vertex in the Bipartite Graph, and ***A*** is the Adjacency Matrix, showing for each (i, j) pair of vertices whether they are adjacent in the Bipartite graph. Adjacent vertices have a 1-valued $A_{ij}$ element and 0-valued otherwise.

A schematic Laplacian matrix is shown in Fig. 4, corresponding to the Modularity Matrix in Fig. 3.

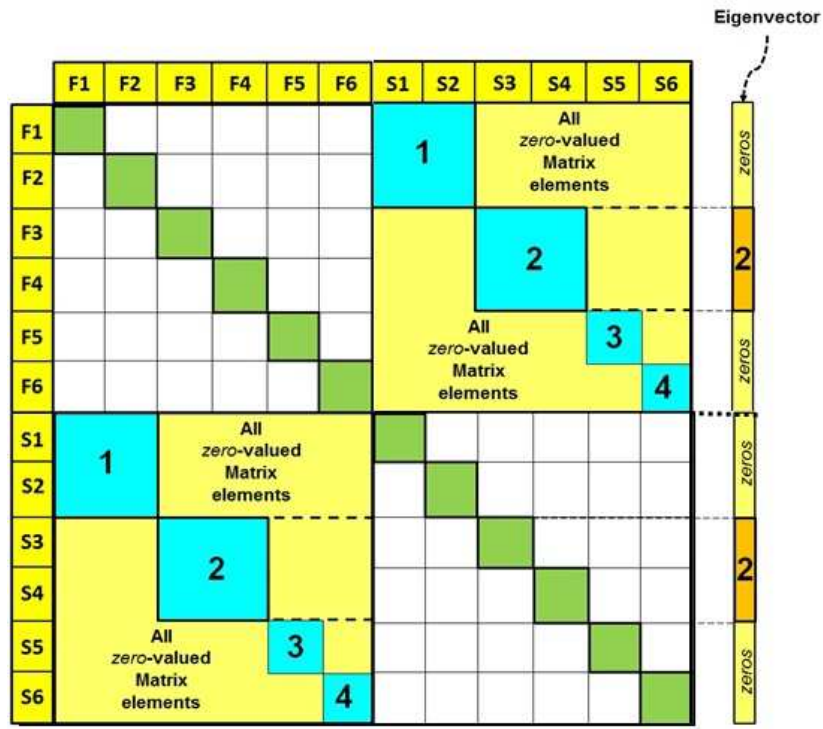

**Figure 4**. – Schematic Laplacian Matrix with eigenvector – It is obtained from the Modularity Matrix in Fig. 3, through the bipartite Structor-Functional graph. The same conventions of Fig. 3 are used here. Laplacian Matrix eigenvectors determine the size of its modules, one eigenvector per module. The Laplacian eigenvector is twice the size of the respective Modularity Matrix eigenvector, seen by comparing this figure with Fig. 3. The Laplacian diagonal (green background) contains the Degree matrix elements. The Laplacian upper-right and lower-left quadrants contain the negative Adjacency matrix, consisting of two copies of the Modularity Matrix reflected around the Laplacian diagonal.

The Degree matrix $D$ is seen in the diagonal of the Laplacian in Fig. 4. The negative adjacency matrix $A$, by equation (1), i.e. two negative copies of the Modularity Matrix (reflecting each other around the Laplacian diagonal) are seen in the upper-right and lower-left quadrants of the Laplacian. The Laplacian eigenvector corresponding to module #2 is double the size of the respective eigenvector of the Modularity matrix in Fig. 3.

In case there are outliers coupling between modules, the Fiedler eigenvector of the Laplacian, corresponding to the next lower eigenvalue (the first non-zero eigenvalue), can be used to formally decouple such a pair of modules. The Fiedler theorem [21] is central to the modularization phase using the Laplacian matrix. Additional details are found in [19].

### 6.5 Summary of Modularization Steps

A series of actions is needed to perform Modularization. These are listed as follows:

a- *Choice of Relevant Matrix* – if it is a Modularity Matrix and one wishes to work with the Laplacian Matrix, obtain the Laplacian through the Bipartite Graph [19];
b- *Matrix Preprocessing* – if it is a Modularity Matrix, symmetrize it and weight it by the appropriate weight values [14];
c- *Calculation of Matrix eigenvectors/eigenvalues* – using a relevant software library;
d- *Choice of suitable Eigenvectors* – according to matrix type: if it is a Modularity Matrix, choose the eigenvectors corresponding to the highest eigenvalues, until the eigenvector modules span the whole matrix; if it is a Laplacian Matrix, choose the eigenvectors whose eigenvalues are zero-valued;
e- *Obtain Modules from eigenvectors* – and calculate the modules sparsity; if sparsity is above threshold and it is a Laplacian Matrix use the Fiedler Vector [19] to split the too sparse module and restart cycle; if sparsity is above threshold and it is a Modularity Matrix, follow the outlier location algorithm in [14], and restart the Software Design Procedure cycle.

## 7. Discussion

We discuss here central issues raised by the "Deconstruction, then Reconstruction" research effort for Software Conceptual Integrity, and related future investigation.

### 7.1 Conceptual Integrity: A Challenging Software Research Journey

The statement by Brooks [3] that "*Conceptual Integrity* is the most important consideration in software system design" is very tempting, as it is not so obvious at first sight. The design principles offered later on by Brooks [4] as a more down to earth interpretation, only add further uncertainty: they do not clarify the idea of Conceptual Integrity and constitute additional independent items to be reformulated and checked for their validity.

It is very satisfying that the Conceptual Integrity idea and two of the offered design principles are ultimately a plausible motivation for the independently developed Linear Software Models [12], [13]. The models were initially proposed solely upon pragmatic size optimization considerations. On the other way round, Linear Software Models provide a formal computational basis for Brooks' ideas. Moreover, Conceptual Integrity and the design principles are related to the four agile-design-rules, originally proposed by Beck [2].

Apparently everything is falling in place. This increases our confidence in this research effort combining the Conceptual Integrity ideas by Brooks, our urgent sense that a theoretical algebraic approach is in demand, and the practical design rule insights by Beck. However, we are not at the end of this challenging journey. We have gained new insights, but much remains to be done, as discussed in the next sub-sections.

### 7.2 Deconstruction, then Reconstruction, of Conceptual Integrity

We have applied the "Deconstruction, then Reconstruction" effort to the ideas and design principles of Frederick Brooks, in order to understand Conceptual Integrity and the eventual relation to its software design principles. Although the ideas and design principles of Brooks were not clear from the beginning, we assumed them to be a reliable source of wisdom, deserving careful analysis.

We were inspired by the "Deconstruction" approach of the French philosopher Jacques Derrida, with the added subsequent "Reconstruction". Similar ideas have previously appeared in the computing literature. Dijkstra aimed by his essay "On the role of scientific thought" [11] *to undo misunderstandings* in the computing realm. On the positive side, Dijkstra expected researchers to gain *renewed understanding* by separation of concerns, thereby enabling a conscious search for useful concepts, in the emerging – here *software* – scientific discipline.

A software engineer should take care not to confuse the – Deconstruction and Reconstruction – philosophical notions with the apparently similar, but specialized technical software concepts of "constructor" and "destructor", found within object-oriented programming languages.

It is beyond this paper's scope to provide a wide enough philosophical background. We just refer the reader to selected literature [9], [10], [34] and recommend Derrida's valuable sources, even though they are no easy reading. Here we focus on a few statements about Derrida's thoughts supporting our "Deconstruction, then Reconstruction" approach. First, Derrida – in the last footnote of its book "*Rogues*" (cited in [10] page 424) – comments that "…Deconstruction does not seek to discredit critique; it in fact constantly relegitimates its necessity…". Second, the translator's preface to "*Of Grammatology*" ([10] page lxix) describes Derrida's approach as: "His text… is the unmaking of a construct. However negative it may sound, deconstruction implies the possibility of *rebuilding*."

### 7.3 Software Conceptual Integrity: Not Anymore a Monolithic Idea

What has been gained by Software Conceptual Integrity "Deconstruction, then Reconstruction"?

We now understand that Software Conceptual Integrity is not anymore a monolithic idea, but it consists of two cyclically interacting phases: Software System Conceptualization and quantitative Modularization of the software system. We also analyzed each of the Brooks' design principles of Propriety and Orthogonality into parts, assigning the obtained parts to the relevant phases, viz. Conceptualization and mainly Modularization.

The most important achievement of "Deconstruction, then Reconstruction" is a refined combination of two concurrent, but contrasting capabilities:

- ***a-*** *clear separation* of Conceptualization from Modularization enabling independent manipulation of each phase by different formal techniques, whenever required;
- ***b-*** *precise transition* between Conceptualization and Modularization in the form of explicit mutual relationships between these two phases of Software Design, embodied e.g. in the Modularity Matrix.

Surprisingly, the Modularization phase is the mainly one motivated by Brooks' design principles, further justifying the "Deconstruction, then Reconstruction" approach.

### 7.4 Elementary Conceptual Kinds for Reconstruction

E*lementary* is meant here as basic for explanation of the theory of software design, in analogy to chemical *elements* in periodic table, or to *elementary* particles in physics. One immediately thinks about open issues of interest, to be discussed here:

- Which are the elementary Conceptual Kinds?
- How many of them should exist?

#### 7.4.1 Conceptualization Reconstruction

The choice of elementary conceptual kinds, in the beginning of section 5, gives a possible answer to the $1^{st}$ issue. The elementary kinds were: a) Domains; b) Ontologies; c) Architectural concepts (i.e. Structors and Functionals); d) Attributes; e) Attribute Ranges of values. All of these are concept kinds, thus relevant to the Conceptualization phase. Their names indicate specialized roles in a software system Conceptualization. This choice is reasonable, but not unique, and probably not definitive.

The second open issue, rephrased here, may help to deal with the previous one:

- What is the optimal (necessary and sufficient) number of elementary conceptual kinds? In particular, what is the optimal number of architectural concepts? Is it just two (Structors and Functionals), three or four?

As a guideline, we avoid proliferation of elementary conceptual kinds by all means. Simplicity is recommendable to facilitate understanding of a software theory. Throughout definitions in sub-sections 5.2 and 5.3 we emphasized the assumption of small numbers.

The number of elementary conceptual kinds issue may remind us the questions of how many design principles or how many agile-design rules are necessary. However, these are different issues.

#### 7.4.2 Modularization Reconstruction

Elementary architectural conceptual kinds – Structors and Functionals – also appear in the matrices of the Modularization phase. This leads to alternative formulations of the second issue above: what is the optimal number of dimensions (or axes) of the Conceptualization space? Are two-dimensional spaces (displayed as 2D-matrices or equivalently bipartite-graphs) sufficient? Since two-dimensional matrices have been successfully used for Modularization, one can argue in favor of just two kinds of elementary architectural concepts. However, again this is not a definitive answer.

There have been other candidates for elementary conceptual kinds in the software literature, such as "properties, requirements, purposes" among others. Let us briefly look at each one of these candidates. Properties seem to be a notion equivalent to attributes. We suggest that requirements and purposes are closely related to functionals.

Requirements are significantly represented in the software engineering literature. As a caveat, here we only refer to functional requirements, since non-functional requirements are not relevant to the current work. Functional requirements can be linked to architectural concepts by means of a so-called Traceability Matrix – usually being a table rather than an algebraic matrix. We argue that a standard Modularity Matrix, with functionals replacing requirements, is necessary and sufficient for modularization analysis.

Purposes have been suggested by Jackson [29], within an ideal mapping design principle, in which each concept is motivated by just one purpose. Problems infringing this principle are an unfulfilled purpose (without a concept), an unmotivated concept (without a purpose), an overloaded concept (with two purposes) and redundant concepts (having the same purpose).

A corresponding matrix modularization analysis in Fig. 5 has columns as concepts **C** and rows as purposes **P**. The ideal mapping is a strictly diagonal matrix. An unfulfilled purpose (empty row), and an unmotivated concept (empty column), are automatically eliminated by modularization. Redundant concepts fit to class inheritance, and overloaded concepts fit a single class providing two different functions, occurring in legitimate software sub-systems.

|    | C1 | C2 |
|----|----|----|
| P1 | 1  | 0  |
| P2 | 0  | 1  |

**Figure 5**. – Schematic Matrix for purposes vs. concepts – Using a Modularity matrix analysis technique, we choose to represent concepts **C** by columns and purposes **P** by rows. This matrix shows the ideal mapping design principle. It is clearly seen as a diagonal matrix, a particular case of the more general block-diagonal matrices, obtained by standard modularization.

Ideal mapping design is correct, but not general enough. Matrix modularization is more generic and has more expressive power. A strictly diagonal system is a particular case of block-diagonal systems, and there is nothing wrong with these more general systems.

Summarizing, we make two claims regarding additional candidates for elementary conceptual kinds. Requirements or purposes:

- a- Neither shift the boundary between Conceptualization and Modularization; i.e. both clearly belong to the Conceptualization phase, which deserves further investigation;
- b- Nor lead to new kinds of modularization analysis.

**7.5 Human Understanding: Deconstruct, then Reconstruct Software Itself**

From the beginning of this paper, the goal of Conceptual Integrity design has been declared to be (in sub-section 1.3) the *human understanding* of the Software System. We have used "Deconstruction, then Reconstruction" to understand Conceptual Integrity, i.e. to grasp its important ideas. Since for every software system we need to achieve a similar goal, viz. to understand the essential concepts of the specific Software System, it is reasonable to apply the same "Deconstruction, then Reconstruction" – or "Analysis, then Synthesis" – approach to the design of each software system itself, as explained and illustrated in the next paragraphs.

Conceptual Deconstruction, to *under-stand* the software system, means to start with the whole software system and gradually decompose it into sub-system concepts, sub-sub-systems, down to the indivisible unit concepts. Reconstruction, to *re*-understand the system, means to compose back from the lowest unit concepts through intermediate sub-systems up to the whole system.

We invoke once more the ATM example (from sub-section 5.6.1). The whole system is represented by the single concept of an Automatic-Teller-Machine. Deconstruction into the next lower abstraction level means to understand the role of each of its biggest sub-systems, which were in our particular example: *bank-account*, *human-machine-interface*, and *communication-and-security*. Continuing deconstruction downwards one sees that *bank-account* is understood by the knowledge of specific types of bank account, decomposing it into *checking-account* and *savings-account*, and so on down to the most elementary concepts, taken as indivisible, e.g. *bank-notes* and *coins*.

Reconstruction involves filling the intermediate sub-systems with all the lacking functionals and respective attributes, until one is satisfied with the system design. It is important to emphasize that Deconstruction and Reconstruction need not be strictly downwards or upwards in the software system hierarchy. The design process, although necessarily iterative, does not need to be unidirectional to fill the still existing gaps.

**7.6 Future Work Needed to Formalize Software Design**

We divide the future work tasks according to the reconstructed Conceptual Integrity embodied in the two phase Software Design Procedure.

The still under development $1^{st}$ Conceptualization Phase needs a non-negligible amount of work to reach a fully formalized mature theory, with a well-based mathematical approach. One probably should start by carefully refining the technique which obtains an Application Ontology for a given software system from the relevant Domain ontologies. The next step could be the choice of a suitable ontology language, hoping to associate a specific algebra with the chosen ontology language. But ontology languages and software tools like Protégé were not formulated with the purpose of software system design. So, it is not just a matter of the most suitable choice among existing options. It may be necessary to rethink the desirable characteristics of the ontology language, having in mind an appropriate algebra for software system design. There may be interesting mathematical alternatives, or an even more radical approach should be taken.

The more mature theory of the 2nd Modularization Phase is based upon a growing self-consistent body of linear algebra knowledge. One has already various significant results (in section 6 of this paper), as follows: for both – Modularity and Laplacian Matrices – there were obtained theorems, and spectral approaches to get modules for the software system, illustrated by case studies; the generation back and forth of the Laplacian Matrix from the Modularity Matrix through the bipartite graph of Structors and Functionals was demonstrated; the equivalence of modules of the Modularity Matrix to the Modularity Lattice has been shown.

Some other desirable results, such as a systematic approach to decouple non-orthogonal modules, within the Modularity Lattice are still lacking. Other potential algebraic results of interest are the subject of current investigation.

### 7.7 Main Contribution

The most important contribution of this paper is the understanding that Conceptual Integrity of a Software System is not a monolithic idea, but a cyclical interaction between software Conceptualization and software Modularization. In practice, it is reflected into concurrent, but contrasting capabilities: <u>*clear separation*</u> of Conceptualization from Modularization, preserving the ability to apply for each of them specific formal manipulation techniques; <u>*precise transition*</u> between Conceptualization and Modularization in the form of explicit mutual relationships between these two phases of Software Design. This is embodied in the Modularity Matrix, enabling fusion of two very different kinds of entities – system concepts and abstract mathematical constructs – as seldom done before.

## Acknowledgments

The author wishes to thank Alessio Plebe from the University of Messina, Italy, and Gonzalo Genova from the University Carlos III of Madrid, Spain, for their incisive and helpful comments on this paper.